\documentclass[twocolumn,english,aps,prc,superscriptaddress,amsmath,amssymb,showpacs]{revtex4}
\usepackage[T1]{fontenc}
\usepackage[latin9]{inputenc}
\usepackage{amsmath}
\usepackage{amssymb}

\usepackage{babel}

\begin{document}

\title{Odd-$J$ Pairing in Nuclei}

\author{L.Zamick}
\author{A. Escuderos}
\affiliation{Department of Physics and Astronomy, Rutgers University, Piscataway,
NJ 08854, USA}

\author{B. Bayman}
\affiliation{University of Minnesota, Minneapolis, USA}

\begin{abstract}
We point out a simplicity that arises when we use an interaction in
which only an energy with odd $J$ is non-zero. The emphasis is on
$J=J_{\text{max}}$ and in particular $J=9^{+}$ in the $g_{9/2}$
shell. It is noted that high overlaps can be deceptive. In many cases
a single set of unitary 9-$j$ coefficients gives either an exact
or a very good approximation to the wave function of a non-degenerate
state. The many degeneracies that occur in these calculations are
discussed and explained.
\end{abstract}

\maketitle

\section{Introduction}

The purpose of this work is to study the properties of a very simple
interaction in a single-$j$-shell model space of neutrons and protons.
A single proton--neutron ($pn$) pair in this space can have a
total angular momentum from $J=0$ to $J=J_\text{max}=2j$. The even-$J$ states
have isospin $T=1$, i.e. they are members of an isotriplet---there are
analog states of two neutrons and of two protons with these even angular
momenta. The odd-$J$ states have isospin $T=0$, i.e. isosinglet---in this
model space, they only are present in the $pn$ system.

There has been much study in journals and in textbooks of the $J=0$,
$T=1$ ``pairing interacton'', i.e. where only $J=0$ two-body matrix
elements are non-zero (and attractive)~\cite{r52,f52,rt53,st63,t93}. Although the pairing
interaction is not realistic, the consequences of these studies has
yielded results whose importance was well beyond expectation. Examples
are seniority classifications, reduced isospin and indeed these studies
can be regarded as precusors to the BCS theory in condensed matter
theory. In turn the BCS theory is important in the nuclear context
for explaining moments of inertia of deformed states. In the $pn$ system,
it is known that not only the $J=0$ two-body matrix element lies low but
also $J=1$ and $J=J_\text{max}=2j$. As a counterpoint to the $J=0$ paring
interaction, we will here consider $J=J_\text{max}$ pairing interaction,
for which $V(2j)$ is non-zero and attractive and all other two-body matrix
elements are zero. Such an interaction only acts between a neutron
and a proton, not between two identical particles. This study will reveal
some surprising results. 

It should be noted that there has been a recent flurry of interest in
$J_\text{odd}$ pairing and in particular the case where $J_{odd}$ is
equal to $J_\text{max}$. In a recent work, C. Qi et al.~\cite{qetal11}
proposed a wave function for $^{96}$Cd in which each of two $pn$
pairs couple to maximum angular momentum, which for the $g_{9/2}$
shell is $J=9$. More generally, they consider the overlap of a wave
function in which one $pn$ pair couples to $J_{1}$ and the other
to $J_{2}$, with basis states of two protons coupling to $J_{p}$
and two neutrons to $J_{n}$. In the latter basis, the effect of the
Pauli principle is much clearer. Once one has antisymmetric wave functions
of two protons, which is achieved by limiting the angular momenta
to even $J_{p}$, and likewise the two neutrons to even $J_{n}$,
one has a wave function which satisfies the Pauli principle. For the
former wave function $[(pn)J_{1}\,(pn)J_{2}]^{I}$, one does have
to antisymmetrize and a priori things look complicated. The authors,
however, make the statement ``The overlap matrix automatically
takes into account the Pauli principle''. Here the overlap matrix
element is 
\begin{equation}
O^{I}(J_{1}J_{2};J_{p}J_{n})=\langle[pn(J_{1})pn(J_{2}){]}^{I}|[pp(J_{p})
nn(J_{n})]^{I}\rangle\,,
\end{equation}
that is, the unitary 9-$j$ symbol, which we will here call $U9$-$j$. 

We here note that work we previously did shows that certain
$U9$-$j$ coefficients form components of the $J=0^{+}$ ground state
wave function of a $J_{\text{odd}}$ pairing interaction. We here
expand on this work. We feel our method is very simple to understand.
We refer to a previous work by E.~Moya de Guerra et al.~\cite{mrzs03}
and explicitly to Eqs. (70) and (74). In that work we describe the
wave functions in a $[(pp)J_{p}\,(nn)J_{n}]$ basis. The Pauli principle
is easily satisfied by constraining $J_{p}$ and $J_{n}$ to be even.
Our previous example was $^{44}$Ti, but the same mathematics holds
for $^{96}$Cd. We previously considered various schematic interactions
as well and the more realistic MBZ interaction taken from experiment~\cite{bmz63},
for which detailed wave funtions were subsequently published in the
archives~\cite{ezb05} (with some modification of the two-body matrix
elements). In this work we use the symbol $I$ for the total angular
momentum of the state and $J$ otherwise.

\section{Calculations}

We now proceed to the calculations. We will be extensively using these
two relations for the $U9$-$j$'s: 
\begin{eqnarray}
\lefteqn{\sum_{J_{13}J_{24}}{\langle(jj)^{J_{a}}(jj)^{J_{b}}|
(jj)^{J_{13}}(jj)^{J_{24}}\rangle^{J}}\times}\nonumber \\
 &  & \times\langle(jj)^{J_{c}}(jj)^{J_{d}}|(jj)^{J_{13}}(jj)^{J_{24}}
 \rangle^{J}=\nonumber \\
 & = & \delta_{a,c}\delta_{b,d}\label{eq:orth}\\
\lefteqn{\sum_{J_{13}J_{24}}{(-1)^{s}\langle(jj)^{J_{12}}(jj)^{J_{34}}|
(jj)^{J_{13}}(jj)^{J_{24}} \rangle^{J}}\times}\nonumber \\
 &  & \times\langle(jj)^{J_{13}}(jj)^{J_{24}}|(jj)^{J_{14}}
 (jj)^{J_{23}}\rangle^{J}=\nonumber \\
 & = & \langle(jj)^{J_{12}}(jj)^{J_{34}}|(jj)^{J_{14}}(jj)^{J_{23}}\rangle^{J}\,,
 \label{eq:u9j-rel2}
\end{eqnarray}
where $s=J_{24}+J_{23}-J_{34}-1$.

Let us first consider the $I=0^{+}$ states in $^{96}$Cd (or $^{44}$Ti).
For most interactions, the diagonalization is a fairly complicated
procedure. However for certain interactions it is much easier. For
example, the interaction used in Ref.~\cite{mrzs03} was one in which
all two-body matrix elements were set equal to zero except for the
$J=1$, $T=0$ two-body matrix element. In such a case, the matrix
element of the secular four-particle Hamiltonian factorizes (the same
result holds for any odd-$J$ interaction). This is the key point.
The basis states are $[J_{p}\, J_{p}]^{I=0}$. We have 
\begin{equation}
H_{J_{p},J_{p'}}=E(J_{\text{odd}})f(J_{p})f(J_{p'})\,,
\end{equation}
 where $f(J_{p})$ is twice the $U9$-$j$ symbol: 
\begin{eqnarray}
f(J_{p}) & = & 2\langle(j^{2})^{J_{p}}(j^{2})^{J_{p}}|(j^{2})^{J_{\text{odd}}}
(j^{2})^{J_{\text{odd}}}\rangle^{I=0}=\nonumber \\
 & = & 2(2J_{p}+1)(2J_{\text{odd}}+1)
 \begin{Bmatrix}
 j & j & J_{p}\\
j & j & J_{p}\\
J_{\text{odd}} & J_{\text{odd}} & 0
\end{Bmatrix}
\end{eqnarray}

If we write the wave function as $\sum X_{J_{p}J_{p}}[J_{p}J_{p}]^{I=0}$
(as in Ref.~\cite{qetal11}), then it was shown in Ref.~\cite{mrzs03}
that $X_{J_{p}J_{p}}$ is proportional to $f(J_{p})$. The other eigenstates
are degenerate and, if $E(J_{\text{odd}})$ is negative, they are
at higher energies. In other words, what we have shown in Ref.~\cite{mrzs03}
is that the wave-function components $X_{J_{p}J_{p}}$ of the lowest
$I=0^{+}$ state are proportional to the overlap factor of Ref.~\cite{qetal11};
alternately, they are equal within a normalization to the $U9$-$j$
coefficients.

The eigenvalue is given by 
\begin{equation}
E(I=0^{+})=E(J_{\text{odd}})|\sum f(J_{p})X_{J_{p}J_{p}}|^{2}
\end{equation}
 Note that our very simple interactions are charge independent. This
means that the lowest (non-degenerate) $I=0^{+}$ state has good isospin,
presumably $T=0$. It is amusing that we can assign the isospin quantum
number to a wave function with $U9$-$j$ coefficients.

In Table~\ref{tab:wf} we present the wave functions for the following
interactions: 
\begin{description}
\item [{CCGI:}] Ref.~\cite{ccgi12}. 
\item [{E(0):}] $V(0)=-2.0000$~MeV; all other matrix elements
are zero. 
\item [{E(9):}] $V(0)=-2.0000$~MeV; all other matrix elements
are zero. 
\item [{E(0,9):}] $V(9)=E(0)=-2.0000$~MeV; all other matrix elements
are zero. 
\item [{E(1):}] $V(1)=-2.0000$~MeV; all other matrix elements
are zero. 
\end{description}

\begin{table}[htb]
 \caption{\label{tab:wf} Wave Functions of the $I=0^{+}$ ground state of $^{96}$Cd
for various interactions.}
\begin{ruledtabular}
\begin{tabular}{cccccc}
 & CCGI  & E(0)  & E(9)  & E(0,9)  & E(1)\\
\hline 
$X_{00}$  & 0.7725  & 0.8563  & 0.6164  & 0.8103  & 0.2903 \\
$X_{22}$  & 0.5280  & 0.1741  & 0.7518  & 0.4814  & 0.5704 \\
$X_{44}$  & 0.2915  & 0.2335  & 0.2385  & 0.2514  & 0.5190 \\
$X_{66}$  & 0.1704  & 0.2807  & 0.0233  & 0.1718  & 0.1586 \\
$X_{88}$  & 0.1020  & 0.3210  & 0.0005  & 0.1831  & $-0.5540$ \\
\end{tabular}
\end{ruledtabular} 
\end{table}

We consider the second column (CCGI) as the realistic interaction
to which the other interactions should be compared. The simplest thing
we can do is give the overlaps of the above interactions with CCGI.
They are respectively 0.9020, 0.9467, 0.9944, and 0.6484.
We find that E(9) gives higher overlap than the much studied E(0)
pairing interaction and a much higher overlap than E(1) . This might
lead one to believe that the idea of $J=9^{+}$ pairing is a valid
concept. But overlaps can be deceiving. We also present E(0,9), where
the only non-vanishing matrix elements are for $J=9^{+}$ and $0^{+}$,
both set to $-2.0000$~MeV. Now the overlap is even higher---0.9944.
This might not be startlingly different than 0.9467, but let us now
look at the energies of the lowest even-$I$ states in Table~\ref{tab:yrast}.
They are given respectively for interactions CCGI, $E(9)$), and E(0,9).
The results of the second column (CCGI) were previously given~\cite{ze12}
and the point was made that the $I=16^{+}$ state is isomeric since
it lies below the lowest $14^{+}$ and $15^{+}$ states. This is in
agreement with experiment~\cite{setal11}.

\begin{table}[htb]
 \caption{\label{tab:yrast} Calculated spectra of yrast even-$I$ states in
$^{96}$Cd for above mentioned interactions.}
\begin{ruledtabular}
\begin{tabular}{cccc}
$I^{\pi}$  & CCGI  & $E(9)$  & E(0,9)\\
\hline 
$0^{+}$  & 0.0000  & 1.0587  & 0.0000 \\
$2^{+}$  & 1.0812  & 1.0589  & 1.2740 \\
$4^{+}$  & 2.1096  & 1.0588  & 1.8584 \\
$6^{+}$  & 2.8883  & 1.0588  & 2.3929 \\
$8^{+}$  & 3.2302  & 1.0571  & 2.5125 \\
$10^{+}$  & 4.8815  & 1.0464  & 3.2142 \\
$12^{+}$  & 5.3394  & 0.9670  & 3.1348 \\
$14^{+}$  & 5.4031  & 0.6570  & 2.8247 \\
$16^{+}$  & 5.2247  & 0.0000  & 2.1678 \\
\end{tabular}
\end{ruledtabular} 
\end{table}

We see that despite the 0.9467 overlap, the even-$I$ spectrum for
Int1 in which only the $J=9^{+}$ matrix element is non-zero is drastically
different than CCGI. First of all, the ground state does not have
$I=0$, rather it has $I=J_{\text{max}}=16$ and indeed the two spectra
seem to have nothing to do with each other.

Let us briefly digress and look at the spectrum for Int1 for its own
sake. It is quite remarkable. The energies of the $I=0,2,4,6$, and
8 states are very close to each other, differing at most by 0.002~MeV
and the $I=10^{+}$ state is 0.012 lower. All six states are essentially
degenerate. Then there is a drop in energy with $I=16^{+}$ becoming
the ground state. Such a strange spectrum and this for an interaction
that gives a 0.9467 overlap with a realistic interaction for the $I=0^{+}$
state.

In the last column of Table~\ref{tab:yrast}, we improve things by
also lowering the $J=0^{+}$ matrix element to the same value as for
$J=9^{+}$, $-2.0000$~MeV. The spectrum is better, with $I=0^{+}$
now the lowest state, but it is far from satisfactory. Even an overlap
exceeding 0.99 does not guarantee overall good results. Clearly all
two-body matrix elements come into play.

As noted above, the eigenfunction of the lowest $I=0^{+}$ state for
the $E(9)$ interaction is $N\langle(j^{2})^{J_{p}}(j^{2})^{J_{n}}|
(j^{2})^{9}(j^{2})^{9}\rangle^{0}$. It can be shown that the normalization 
factor is $\sqrt{2}$.

For the $I=1^{+}$ states with the $E(9)$ interaction, the secular
matrix is also separable. This is not true for other values of $J_{\text{odd}}$.
If we were to replace $I=0$ by $I=1$ in the above expression, all
the $U9$-$j$ coefficients would vanish. We must make a different
choice. The eigenfunction of the lowest $I=1^{+}$ state is then given
by a single set of $U9$-$j$ coefficients: $2\langle(j^{2})^{J_{p}}(j^{2})^{J_{n}}|
(j^{2})^{9}(j^{2})^{8}\rangle^{I=1}$. This state has isospin $T=1$. 
Indeed all $I=1^{+}$ states in this model space have isospin $T=1$. 
The other four $I=1^{+}$ states are degenerate at a higher energy.

For states with $I=2$ or higher, the secular matrix is no longer
separable---rather it is a sum of separable terms. The eigenvalue
equation is 
\begin{eqnarray}
\lefteqn{4\sum_{J_{x}}\langle(j^{2})^{J_{p}}(j^{2})^{J_{n}}|
(j^{2})^{9}(j^{2})^{J_{x}}\rangle^{I}\times}\nonumber \\
 &  & \times\sum_{J_{p'}J_{n'}}\langle(j^{2})^{J_{p'}}(j^{2})^{J_{n'}}|
 (j^{2})^{9}(j^{2})^{J_{x}}\rangle^{I}D(J_{p'},J_{n'})=\nonumber \\
 & = & \lambda D(J_{p},J_{n})
\label{eq:i2ev}
\end{eqnarray}
For $I=2^{+}$ there are two terms corresponding to $J_{x}=7$ and
9; for $I=3^{+}$ the values are $J_{x}=6$ and 8, etc.

Despite the complexity of the above equation, there are some surprising
results. The eigenfunction components of the lowest $2^{+}$ state
are numerically extraordinarily close to the single $U9$-$j$ symbols
$\sqrt{2}\langle(j^{2})^{J_{p}}(j^{2})^{J_{n}}|(j^{2})^{9}(j^{2})^{9}\rangle^{I=2}$.
Furthermore, the next $2^{+}$ state has also components exceedingly
close to $2\langle(j^{2})^{J_{p}}(j^{2})^{J_{n}}|(j^{2})^{9}(j^{2})^{7}\rangle^{I=2}$.
This is by no means obvious because, as mentioned above, the interaction
involves a sum of two separable terms corresponding to $J_{x}=7$
and 9. We can explain this result by performing the sum over even
$J_{p}$ and even $J_{n}$. We first note schematically 
\begin{eqnarray}
\lefteqn{4\sum_{\text{even }J_{p}J_{n}}=\sum(1+(-1)^{J_{p}})(1+(-1)^{J_{n}})=}\nonumber \\
 & = & \sum+\sum(-1)^{J_{p}}+\sum(-1)^{J_{n}}+\sum(-1)^{J_{p}+J_{n}}\label{eq:sch}
\end{eqnarray}

The first term vanishes because of Eq.~(\ref{eq:orth}). In the last
term one of the $U9$-$j$'s has two rows that are the same, which
means that the only non-vanishing terms in the sum have ($J_{p}+J_{n}$)
even. Thus, the last term is the same as the first term---zero. The
two middle terms are the same, so we get 
\begin{eqnarray}
\sum_{\text{even }J_{p}J_{n}} & = & \frac{1}{2}\sum(-1)^{J_{p}}\langle(j^{2})^{J_{p}}
(j^{2})^{J_{n}}|(j^{2})^{9}|(j^{2})^{9}\rangle^{I=2}\times\nonumber \\
 &  & \times\langle(j^{2})^{J_{p}}(j^{2})^{J_{n}}|(j^{2})^{9}(j^{2})^{7}\rangle^{I=2}=\nonumber \\
 & = & -\frac{1}{2}\langle(j^{2})^{9}(j^{2})^{9}|(j^{2})^{9}(j^{2})^{7}\rangle^{I=2}\,.
 \label{eq:i2}
\end{eqnarray}
We obtain the above by using two orthogonality relations for $9j$-symbols
as shown e.g. in \textit{Nuclear Shell Theory} (p.~516) by de-Shalit
and Talmi~\cite{st63}. We call the right-hand side of Eq.~(\ref{eq:i2})
the overlap.

Using similar arguments, one can show that the normalization for the
$[9\,,9]$, $N$, is such that 
\begin{eqnarray}
N^{-2} & = & \frac{1}{2}-\frac{1}{2}\langle(j^{2})^{9}(j^{2})^{9}|
(j^{2})^{9}(j^{2})^{9}\rangle^{I=2}=\label{eq:n2-99}\\
 & = & \frac{1}{2}-\frac{1}{2}0.00001209813=0.499993950935\nonumber 
\end{eqnarray}

For the $[9\,,7]$ case, we obtain 
\begin{eqnarray}
N^{-2} & = & \frac{1}{4}-\frac{1}{2}\langle(j^{2})^{9}(j^{2})^{7}|
(j^{2})^{9}(j^{2})^{7}\rangle^{I=2}=\label{eq:n2-97}\\
 & = & \frac{1}{4}+\frac{1}{2}0.00075253477=0.250376267385\nonumber 
\end{eqnarray}

To get this latter result, we use the following relationship 
\begin{equation}
\sum(-1)^{(J_{p}+J_{n})}\left|\langle(j^{2})^{9}(j^{2})^{7}|
(j^{2})^{J_{p}}(j^{2})^{J_{n}}\rangle^{I=2}\right|^{2}=0
\end{equation}

Therefore, we obtain that the overlap is exceedingly small for the
$g_{9/2}$ shell. From Eqs.~(\ref{eq:n2-99}) and (\ref{eq:n2-97}),
we find that the normalizations are 1.414222 and 1.998497, the latter
slightly smaller than 2. The overlap in Eq.~(\ref{eq:i2}) is 0.00009113
and, if we include the normalization factors, we get 0.00025756.

In lower shells the deviations are larger. For example, in the $d_{5/2}$
shell, we replace $[9\,,9]$ by $[5\,,5]$ and $[9\,,7]$ by $[5\,,3]$.
The overlap is now 0.0107, small but not zero. The norms are no longer
$\sqrt{2}$ and 2, but rather 1.4161 and 1.9204. With normalized states
the overlap is 0.0290. The corresponding numbers in the $f_{7/2}$
shell are 0.001037, 1.41434, 1.9875 and 0.002916. The overlap is here
smaller than in the $d_{5/2}$ shell, but it is not zero. As one can
see above in the $g_{9/2}$ shell, the overlap is an order of magnitute
smaller than in the $f_{7/2}$ shell. We can speculate that the overlap
might vanish in the infinite-$j$ limit.

We can see in Table~\ref{tab:j2} that the results for matrix diagonalization
for both $I=2^{+}$ states yield wave function components which are
very close to the normalized $U9$-$j$ coefficients. In fact, they
are so close that one could wonder if they are exactly the same. But
they are not. As seen in Eq.~(\ref{eq:i2}), the two $U9$-$j$ sets
corresponding to $[9\,,9]$ and $[9\,,7]$ are very nearly orthogonal,
but not quite.

The above $I=2$ states have isospin $T=0$. The $E(9)$ interaction also yields
a $T=1$ non-degenerate state with components $2 \langle (jj)^{9} (jj)^{8}|
(jj)^{J_{p}} (jj)^{J_{n}} \rangle^{2}$. This is a pure state---it does
not mix with any other $T=1$ state. This is because there is only one
way of forming an $I=2$, $T=1$ state from $U9$-$j$ symbols, i.e only one possible
$J_{x}$.

For $I=3$, $T=0$ there is also a pure state $2\langle (jj)^{9} (jj)^{7}|
(jj)^{J_{p}} (jj)^{J_{n}} \rangle^{3}$. This wave function changes
sign under the interchange of $J_{p}$ and $J_{n}$. It cannot admix
with a state with $J_{x}=8$ for which there is no change of sign
when $J_{p}$ and $J_{n}$ are interchanged.

\begin{table}[htb]
\caption{\label{tab:j2} Comparison for the first two $I=2^{+}$ states of
the matrix diagonalization with the $E(9)$ interaction and with normalized
$U9$-$j$ components. We give the energy in MeV in the second row.}
\begin{ruledtabular}
\begin{tabular}{ccccc}
$[J_{p}\,,J_{n}]$  & $E(9)$  & $U9$-$j$  & $E(9)$  & $U9$-$j$ \\
 & 1.069  &  & 3.0558  & \\
\hline 
$[0\,,2]$  & 0.5334  & 0.5338  & 0.1349  & 0.1351 \\
$[2\,,2]$  & $-0.4707$  & $-0.4708$  & 0.5569  & 0.5567 \\
$[2\,,4]$  & 0.3035  & 0.3035  & 0.3188  & 0.3189 \\
$[4\,,4]$  & $-0.1388$  & $-0.1390$  & 0.6300  & 0.6299 \\
$[4\,,6]$  & 0.0531  & 0.0531  & 0.1320  & 0.1320 \\
$[6\,,6]$  & $-0.0137$  & $-0.0138$  & 0.1350  & 0.1350 \\
$[6\,,8]$  & 0.0025  & 0.0025  & 0.0114  & 0.0114 \\
$[8\,,8]$  & $-0.0003$  & $-0.0003$  & 0.0052  & 0.0052 \\
\end{tabular}
\end{ruledtabular} 
\end{table}

It turns out that all the other lowest even-$J$ states have eigenfunctions
close although not exactly equal to $\sqrt{2}\langle(j^{2})^{J_{p}}(j^{2})^{J_{n}}|(j^{2})^{9}(j^{2})^{9}\rangle^{I}$.
In Table~\ref{tab:9j} we compare, as an example, the wave function
of the $J=8^{+}$ state. In the second column, we give the single
$U9$-$j$ symbols (normalized) and in the third column we give results
of diagonalizing the $E(9)$ interaction. Since the coefficient $[J_{p},J_{n}]$
is the same as $[J_{n},J_{p}]$, we list only one of them. The overlap
of the two wave funtions, $\langle\psi_{1},\psi_{2}\rangle$, is 0.9944. 

\begin{table}[htb]
\caption{\label{tab:9j} Comparing the wave functions of a single $U9$-$j$
symbol with $J_{x}=9$ with a full diagonalization of $E(9)$ for
the lowest $I=8^{+}$ state in $^{96}$Cd.}
\begin{ruledtabular}
\begin{tabular}{ccc}
$[J_{p}\,,J_{n}]$  & $U9$-$j$  & $E(9)$ \\
\hline 
$[0\,,8]$  & 0.0630  & 0.0644 \\
$[2\,,6]$  & 0.4299  & 0.4271 \\
$[2\,,8]$  & $-0.0522$  & $-0.0513$ \\
$[4\,,4]$  & 0.7444  & 0.7456 \\
$[4\,,6]$  & $-0.1803$  & $-0.1729$ \\
$[4\,,8]$  & 0.0256  & 0.0280 \\
$[6\,,6]$  & 0.0521  & 0.0657 \\
$[6\,,8]$  & $-0.0076$  & $-0.0012$ \\
$[8\,,8]$  & 0.0011  & 0.0047 \\
\end{tabular}
\end{ruledtabular} 
\end{table}


\section{Degeneracies}

With this $E(9)$ interaction, we get several degenerate states with 
an absolute energy zero. In some detail, for $I=0$ there are five 
states, three with isospin $T=0$ and two with $T=2$. There is one 
non-degenerate state at an energy $2V(9)$ ($V(9)$ is negative). The 
other four $I=0$ states have zero energy.
For $I=1$ all states have isospin $T=1$. There is a single non-degenerate 
state at $V(9)$, the other three have zero energy.
For $I=2$ there are twelve states---six have $T=0$, four have $T=1$ 
and two have $T=2$. There are two non-degenerate $T=0$ states with 
approximate energies $2V(9)$ and $V(9)$ respectively, and one 
non-degenerate $T=1$ state with energy $V(9)$. The other nine states 
have zero energy.
To understand this, take a wave function 
\begin{equation*}
\Psi^{\alpha} = \sum{C^{\alpha} (J_p, J_n) [J_p J_n]^{I}}
\end{equation*}
and the corresponding energies $E^{\alpha} = \langle \Psi^{\alpha} 
H \Psi^{\alpha} \rangle$. Consider the sum $\sum_{\alpha}{E^{\alpha}}$.
We have
\begin{equation}
\sum_{\alpha}{C^{\alpha} (J_p, J_n) C^{\alpha} (J_{p'}, J_{n'})} = 
\delta_{J_p,J_{p'}} \delta_{J_n,J_{n'}}
\end{equation}
Thus
\begin{eqnarray}
\sum_{\alpha}{E^{\alpha}} & = & \sum_{J_p J_n}{\langle [J_p J_n]^{I} 
 H [J_p J_n]^{I} \rangle} = \\
 & = & 4V(9) \mathop{\sum_{J_p J_n}}_{\text{even}}{\sum_{J_A}{\left| 
 (jj)^{J_p} (jj)^{J_n} | (jj)^{9} (jj)^{J_A} \right|^{2}}} \nonumber
\end{eqnarray}
This expression does not depend on the detailed wave functions. 
Using the properties of $U9$-$j$'s, we can show that $\sum_{\alpha}
{E^{\alpha}} = 2V(9)$ for $I=0$, $V(9)$ for $I=1$, and $4V(9)$ for $I=2$. 
But we can alternately show, using the explicit wave functions, that for 
$I=0$ the energy of the lowest state is $2V(9)$. Hence, all the other 
states must have zero energy. A similar story for $I=1$.
The $I=2$ state is a bit more complicated because of the coupling between 
two states, however small it is. Still one can work it through and see 
that the $4V(0)$ energy is exhausted by the two $T=0$ and the one $T=1$ 
non-degenerate states.

\section{Closing remarks}

In closing, we note that the subject of $J_{\text{max}}$ pairing
is currently a very active field. Besides the work of Qi et al.~\cite{qetal11},
there are related works by Zerguine and Van Isacker~\cite{zi11},
Cederwall et al.~\cite{cetal11} and Xu et al.~\cite{xetal12}.
The topic of $J$-pairing interactions has also been addressed by
Zhao and Arima~\cite{za05}. In this work we expand on our 2003 work~\cite{mrzs03}
by renoting that the Hamiltonian matrix for a $2p$-$2n$ system for
$I=0^{+}$ states in a single $j$-shell is separable for a simple
interaction which is non-zero only for a single odd angular momentum.
This leads to an eigenfunction with components proportional to a single
set of unitary 9-$j$ symbols. We apply this to the $J=J_{\text{max}}$
interaction. The single set of $U9$-$j$ components form the eigenfunction
not only for the lowest $I=0^{+}$ state, but also of the lowest $I=1^{+}$
state and, to a surprisingly excellent approximation, for the lowest
two $I=2^{+}$ states. A single set of $U9$-$j$ coefficients yields
a good approximation for the higher yrast even-$I$ states. We use
the $J=J_{\text{max}}$ interaction to confirm the observation of
Ref.~\cite{qetal11} that the resulting wave function has a fairly
high overlap with that of a realisitc interaction, although we note
the energies are not realistic. We also note that this wave function
for the lowest $I=0^{+}$, $1^{+}$, and $2^{+}$ states has good
isospin. We have other examples too: we have found a quantum number
$J_{x}$ (see Eq.~(\ref{eq:i2ev})) which can, either exactly or approximately,
help classify some of the states.

Our cautionary remarks are for the topic of pairing in general, but
are not intended specifically for any of the works mentioned in this
paper.

\end{document}